\documentclass[twoside]{article}
\usepackage{fleqn,espcrc2,epsfig}

\newcommand{\AmS}{{\protect\the\textfont2
  A\kern-.1667em\lower.5ex\hbox{M}\kern-.125emS}}

\title{Composite Reweighting the Glasgow Method for Finite Density 
QCD}

\author{ P. R. Crompton
\address{Dept. of Physics and Astronomy, University of Glasgow, G12 8QQ, Scotland, UK.}}

\begin{document}
\pagestyle{empty}
\setcounter{topnumber}{1}

\begin{abstract}
%----------------------------------------------------------------
{The reweighting 
scheme developed in Glasgow to circumvent the lattice action becoming complex 
at finite density suffers from a 
pathological onset transition thought to be due to the 
reweighting. We present a new reweighting scheme based on this approach in 
which we combine ensembles to alleviate the sampling bias we 
identify in the polynomial coefficients 
of the fugacity expansion.}

\end{abstract}

\maketitle

\section{Introduction}
Lattice simulations of dense matter are important for the understanding of future 
heavy ion collider experiments as well as the equation of state of neutron 
stars \cite{20}\cite{20a}. The 
main obstacle to the numerical evaluation of QCD in this regime 
is that the lattice action becomes 
complex with the inclusion of the chemical potential $\mu$, 
prohibiting naive probabilistic Monte Carlo methods. Recent attempts to evaluate QCD at 
finite baryon density have involved evaluating canonical ensembles with a given 
number of background quark sources \cite{21}\cite{22}, and the evaluation of QCD-like models in 
which the 
lattice action is real at finite $\mu$ \cite{23}\cite{24}. 
An earlier reweighting approach was developed at Glasgow, in which 
the Grand Canonical Partition function is evaluated semi-analytically.   
%-------------------------------------------------------------------------------
\section{The Glasgow Method}

The $\mu$ dependence of the lattice action is made 
analytic in the Glasgow reweighting method 
through the formulation of a characteristic polynomial in the 
fugacity variable $z = {\rm{exp}}(\mu/T)$ \cite{1}.  
The fermion matrix $M$ (defined with Kogut-Susskind fermions \cite{2}) 
is re-expressed in terms of the matrices which contain only links between lattice sites 
in the spatial directions $G$, and forward and backward in the time direction 
$V$ and $V^{\dagger}$. which allows the definition of the propagator matrix 
$P$ \cite{3}.

\begin{equation}
2iM = 2im + G + V e^{\mu} + V^{\dagger}e^{-\mu} 
\end{equation}

\begin{equation}
P = \left( \begin{array}{cc} 
 -( G + 2im ) & 1  \\
-1	       & 0   \end{array}\right) V
\label{P}
\end{equation}

The propagator matrix is then used to 
re-express ${\rm{det}} M$ as a characteristic polynomial in the variable 
$e^{-\mu}$,
 
\begin{eqnarray}
{\rm{det}} M  & = & {\rm{det}}  ( G + 2im + 
V^{\dagger}e^{-\mu} + Ve^{\mu} ) \\
& = & e^{n_{c}n_{s}^{3}n_{t}\mu} \,\,{\rm{det}}  ( P - e^{-\mu} )\\
& = & e^{n_{c}n_{s}^{3}n_{t}\mu}\sum_{n=0}^{2n_{c}n_{s}^{3}n_{t}}
c_{n}e^{-n\mu} 
\label{cn}
\end{eqnarray}
where $n_{s}^{3}n_{t}$ is the lattice volume, $n_{c}$ the number 
of colours, and the expansion coefficients $c_{n}$, so defined, 
are functionals of the lattice gauge fields. 
Since $V$ is an overall factor of $P$ the expansion can be 
further simplified through the symmetry $Z_{n_{t}}$ associated with 
performing a unitary transformation on $P$ by multiplying the 
timelinks by $e^{2\pi ij / n_{t}}$, where $j$ is an integer. 
This then allows relation of the expansion coefficients of the characteristic 
polynomial (where $n_{t} = 1/T$) to the canonical 
partition functions $Z_{n}$ and the Grand Canonical Partition function 
$Z(\mu)$. 
As Lee and Yang showed with an Ising ferromagnetic system, 
in the thermodynamic limit a 
phase transition occurs wherever a zero of the fugacity polynomial 
approaches the real 
axis in the complex-$z$ plane \cite{4}. 
The zeros $\alpha_{n}$ are determined numerically by rootfinding the 
reweighted polynomial expansion coefficients. 

\begin{eqnarray}
Z(\mu) & = & \int DU \,\, {\rm{det}} M(\mu) \,\, e^{-S_{g}} \\ 
& = & \sum_{n} \,\, Z_{n} \,\, e^{n\mu/T} \\
\frac{Z(\mu)}{Z(\mu_{o})} & = & \frac {\int DU \,\, {\displaystyle{\frac{ {\rm{det}} 
M(\mu) }{ {\rm{det}} 
M(\mu_{o}) }}}
\,\,\, {\rm{det}} M(\mu_{o}) \,\, e^{-S_{g}}} {\int DU 
\,\, {\rm{det}} M(\mu_{o}) \,\, e^{-S_{g}}} 
\\ 
\label{norm}
 & = & \left\langle {\frac{{\rm{det}} M(\mu)}{{\rm{det}} M(\mu_{o})}} 
\right\rangle_{\mu_{o}}\\
& \propto & e^{-n_{c}n_{s}^{3}n_{t}\mu} \,\,
\prod_{n=1}^{n_{c}n_{s}^{3}} \,\,( e^{n_{t} \mu} - \alpha_{n} ) 
\end{eqnarray}

Reweighting introduces normalisation by the ensemble generated at $\mu_{o}$ into 
Eqn.(\ref{norm}), which leaves the 
analytic determination of the critical points unaffected. 
The advantage of the scheme is that, 
even though the fermionic action is in general complex 
in SU(3), an ensemble can be generated at $\mu_{o} = 0$. 
However, it has been shown  that the unphysical onset transition at 
$\mu = \frac{1}{2}m_{\pi}$ of quenched measurements \cite{5}, 
persists despite the inclusion of 
dynamical quarks. From which it is concluded 
that the Monte Carlo sampling is ineffective, because the $\mu_{o} = 0$ ensemble has little 
overlap with the physically relevant region \cite{6}\cite{6a}. 
Parallels can also be drawn between this 
pathology and the sign 
problem of the related reweighting of the Hubbard model \cite{7}.

\section{SU(2) at Intermediate Coupling}

\subsection{Composite Reweighting}
%-------------------------------------------------------------------------------
For SU(2) we can vary the value of $\mu_{o}$ we use to generate ensembles (as the group is 
pseudoreal for quarks in the fundamental representation \cite{7a}). 
This then allowed us to investigate the reliability of the ensemble-averaging 
of the polynomial expansion coefficients of Eqn.(\ref{cn}), 
which is defined through the fugacity expansion. 

\begin{eqnarray}
\frac {Z_{n}} {Z(\mu_{o})} & = & \frac {\int DU {\displaystyle{\frac {c_{n}} { {\rm{det}} 
M(\mu_{o}) }}}
\,\, {\rm{det}} M(\mu_{o}) \,\, e^{-S_{g}}} {\int DU 
\,\, {\rm{det}} M(\mu_{o}) \,\, e^{-S_{g}}} \\
\label{aver}
& = & \left\langle {\frac{{c_{n}}}{{\rm{det}} M(\mu_{o})}}
 \right\rangle_{\mu_{o}} 
\end{eqnarray}

\begin{table}
\begin{center}
\begin{tabular}{|c|c|c|c||}		\hline
$\mu_{o} $ & Re $\eta_{1}$ & Im $\eta_{1}$ & $\mu ( {\rm {max}} \chi_{n} )$

\\ \cline{1-1} \cline{2-2} \cline{3-3} \hline

0.3	& 0.411(0.001) & 0.116(0.001) & 0.41(0.01)  \\ \hline

0.5	& 0.830(0.002) & 0.167(0.096) & 0.83(0.01)  \\ \hline

0.7	& 0.523(0.003) & 0.134(0.001) & 0.52(0.01)  \\ \hline
	
0.8	& 0.822(0.028) & 0.154(0.082) & 0.82(0.01)  \\ \hline

0.9	& 0.546(0.067) & 0.153(0.051) & 0.55(0.01)  \\ \hline

1.0	& 0.434(0.039) & 0.091(0.039) & 0.43(0.01)  \\ \hline

1.1	& 0.461(0.011) & 0.064(0.030) & 0.46(0.01)  \\ \hline

$\#. \, \rm{Ens.} $ & - & - & -

\\ \cline{1-1} \cline{2-2} \cline{3-3} \hline

1	& 0.546(0.067) & 0.153(0.051) & 0.55(0.01)  \\ \hline

3	& 0.467(0.008) & 0.012(0.007) & 0.47(0.01)  \\ \hline

5	& 0.453(0.008) & 0.011(0.007) & 0.45(0.01)  \\ \hline
	
7	& 0.477(0.001) & 0.006(0.005) & 0.48(0.01)  \\ \hline

\end{tabular}	
\caption{Dependence of $\eta_{1}$ for SU(2) at intermediate coupling 
on the $\mu_{o}$ (upper), 
and number of ensembles included in the composite reweighting (lower).}
\end{center}
\end{table} 
These coefficients are evaluated recursively from the propagator matrix $P$, 
and the denominator ${\rm{det}}M(\mu_{o})$ by evaluating Eqn.(\ref{cn}) at 
$\mu = \mu_{o}$.  
Where the ratio on the lefthand side of Eqn.(\ref{aver})
differs from one, we found from our measurements that the reliability of 
the ensemble-averaging is strongly affected. In turn this 
affected our determination of the Lee Yang zeros through rootfinding, and our 
measurement of related thermodynamic observables evaluated from 
the expansion coefficients. 
We monitored the effect by measuring the ratio $W^{21}(n)$, 
of the expansion coefficients from ensembles 
generated at two different values of $\mu_{o}$. 

\begin{eqnarray}
W^{21}(n)  & = & \left\langle {\frac{c_{n}}{{\rm{det}} M(\mu_{2})}} 
 \right\rangle_{\mu_{2}}
\left\langle {\frac{c_{n}}{{\rm{det}} M(\mu_{1})}} 
\right\rangle_{\mu_{1}}^{-1}
\\ & = & \frac{Z( \mu_{1} )}{ Z( \mu_{2} )} 
\end{eqnarray}

Although this ratio should be independent of $n$, 
our measurements indicate that 
ensemble-averaging selectively affects the determination of the expansion. 
Only a small range of $n$ can be reliably determined 
from an ensemble generated at a given value of $\mu_{o}$. However, since this 
small range centers on different $n$ as $\mu_{o}$ is varied, 
coefficients from a covering of several ensembles (generated at different 
values of $\mu_{o}$) can be combined to alleviate the bias inherent in 
the reweighting. 
The reliably ensemble-averaged range of coefficients 
from the ensemble generated at $\mu_{2}$ (multiplied by $W^{21}$) is used to 
replace the corresponding range of the expansion of the ensemble generated at 
$\mu_{1}$. As the number of covering ensembles is 
increased
these new composite reweighted coefficients, which replace the ill-determined 
coefficients, thus converge to accurate values.

\subsection{Results}

Measurements of the chiral $\langle \overline{\psi}\psi \rangle$ and 
diquark condensates 
$\langle \psi\psi \rangle$ at strong \cite{7b}
and intermediate coupling 
\cite{7c}\cite{8}\cite{8a}\cite{8b}, indicate  
at least two distinct finite density regimes. 
The first, a low density vacuum regime in which 
$\langle \overline{\psi}\psi \rangle \neq 0$ $\langle \psi\psi \rangle =0$, is 
believed to be separated from a thermodynamic regime, where 
$\langle \overline{\psi}\psi \rangle = 0$ 
$\langle \psi\psi \rangle \neq 0$, by a 
transition at $\mu = \frac{1}{2}m_{\pi}$. The $U(2)$ 
symmetry at 
$m, \mu =0$ allows a rotation between the two condensates and so 
the superfluid diquark can be used as an indicator of the spontaneous breaking 
of the staggered fermion chiral symmetry remnant $U(1)_{A}$ 
(in addition to $U(1)_{V}$). 
\begin{figure}
\epsfig{file=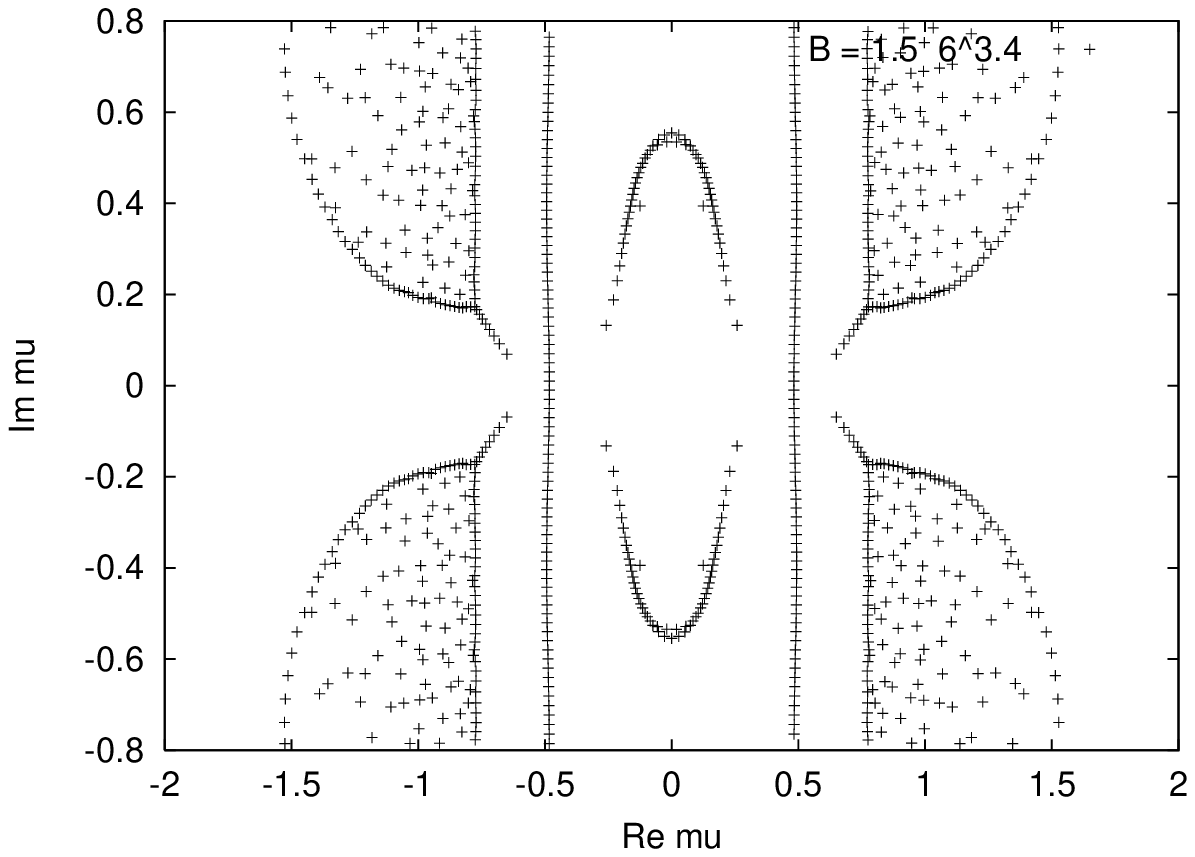, height=0.28\textheight}
\vspace{-1.0cm}
\caption{Lee Yang zeros evaluated in the complex 
$\mu$ plane for SU(2) at intermediate 
coupling from seven composite reweighted ensembles. }
%\label{fig:results}
\vspace{1.2cm}
\epsfig{file=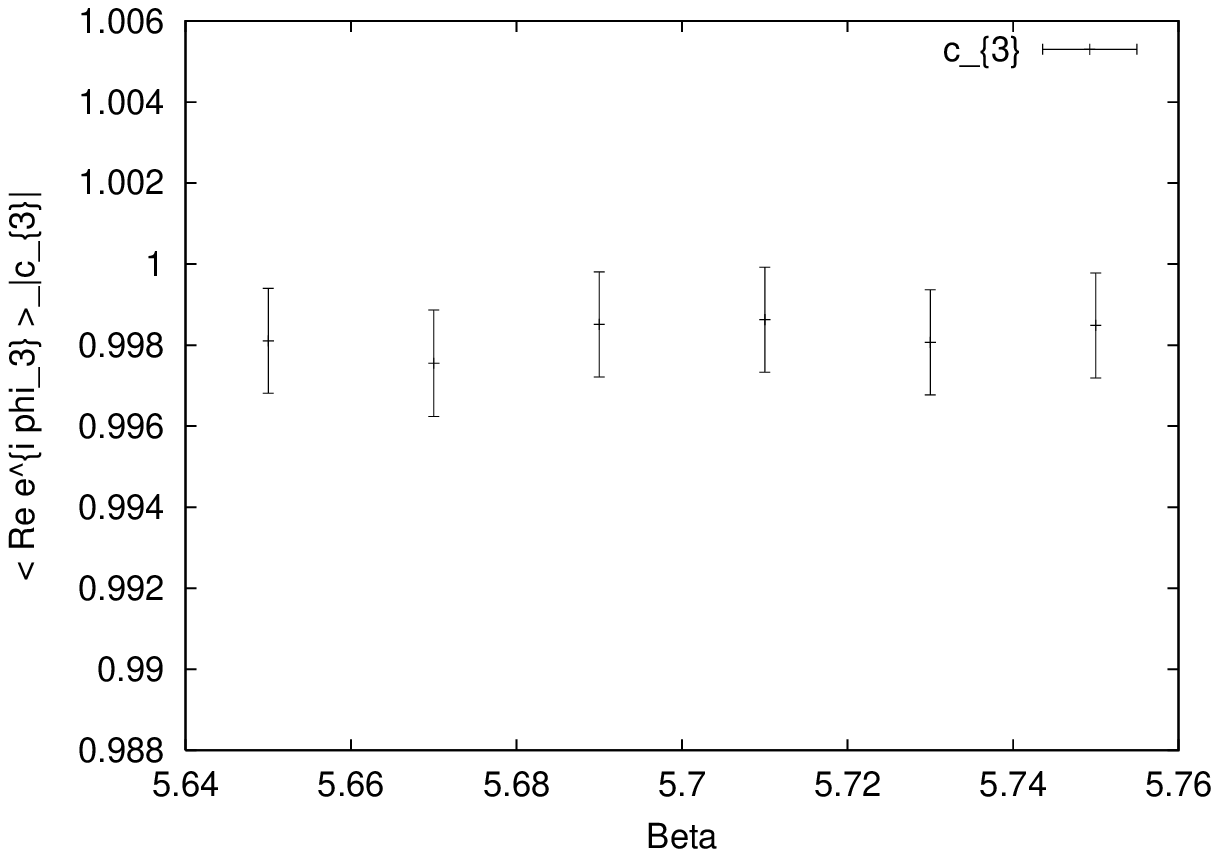, height=0.28\textheight}
\vspace{-1.0cm}
\caption{ $\langle \, Re \, 
(e^{i\phi_{n}})  \rangle_{|c_{n}|}$ the ensemble-averaged real part of the phase of 
the polynomial expansion coefficient $c_{n}$ (for $n=3$), which we use as a Monte 
Carlo measure for static SU(3) at intermediate coupling.} 
%\label{fig:results}
\end{figure}

Our measurements of the zeros of the Grand Canonical Partition function 
identify this transition as we tabulate in Table 1. 
By evaluating the Lee Yang zeros $\alpha_{n}$ in the complex $\mu$ plane 
 we 
can readily identify the value of $\mu$ at the transition. In 
the thermodynamic limit ${\rm{Im}}\, \eta_{1} \rightarrow 0$ and so 
${\rm{Re}}\, \eta_{1}$ corresponds to $\mu_{c}$.
\begin{equation}
\eta_{n} = T \, {\rm{ln}} \, \alpha_{n} 
\end{equation}
Before our composite 
reweighting procedure is applied, $\eta_{1}$ 
is inconsistent and strongly dependent on the value of 
$\mu_{o}$ used in the 
reweighting. However the jacknife error estimates with our new scheme 
indicate that the zeros can be consistently determined where we alleviate 
the inaccuracies caused by reweighting. As we increase the number of 
ensembles generated at succesive values of $\mu_{o}$ used to cover the 
expansion, ${\rm{Re}}\, \eta_{1}$ becomes consistent. Also 
${\rm{Im}}\, \eta_{1} \rightarrow 0$, from which we are able to more 
confidently associate ${\rm{Re}}\, \eta_{1}$ with $\mu_{c}$. 
Our measurements at 
$\beta =1.5, m= 0.05$ on a $6^{3}4$ lattice indicate a transition at 
$\mu = 0.48$, which is in approximate agreement with existing measurements 
\cite{7a}\cite{su2}\cite{su2a}\cite{su2b}\cite{su2c}. 
 The prominent peaking of our further 
measurements of the 
quark number density susceptibility from the composite weighted expansion 
coefficients, we believe, indicates that this transition is 
first order.

\section{Static SU(3) at Intermediate Coupling}
Having developed our new approach in SU(2), we now evaluate the 
ensemble-averaged polynomial 
expansion coefficients of static SU(3) \cite{9}, 
from canonical ensembles generated 
with different numbers of background quark sources. 
In this way we are able to again systematically vary the localised region in 
which the ensemble-averaging is reliable, and combine coefficients from 
different ensembles.

\begin{eqnarray}
\langle {\mathcal{O}} \rangle_{Z_{n}} & = & 
\langle {\mathcal{O}} \rangle_{c_{n}} \\ 
& = & 
\frac {\int DU \,\,\, {\mathcal{O}} \,\,\,c_{n} \,e^{-S_{g}}}{\int DU 
\,c_{n} \, e^{-S_{g}}} \\   & = & 
\frac{\langle \, {\mathcal{O}} \, e^{i\phi_{n}} \, 
\rangle_{|c_{n}|}}{ 
\langle \, e^{i\phi_{n}} \, \rangle_{|c_{n}|}}  
\end{eqnarray}

Our implementation of static SU(3) involves setting $G=0$ in Eqn.(\ref{P}) 
which differs 
slightly from existing schemes \cite{9}\cite{9a}\cite{9b} 
in that we additionally incorporate the 
relativistic effect of antiquarks. The motivation for choosing this model 
over full SU(3) is purely the computational expediency that refreshing 
$c_{n}$ in one dimension alone affords the Monte Carlo sampling.  
Although the expansion coefficients 
are in general complex, our measurements of the ensemble-averaged real part of 
the phase of our measure indicate that the sign problem of the measure 
is negligible, Fig.(2). 
As $n$ moves away from the index of the expansion coefficient 
used as the Monte Carlo measure, however, the ensemble averaging of the 
expansion coefficients becomes less 
effective, as we monitor in Fig.(3). For the two ensembles generated with the 
measures $c_{75}$ and $c_{51}$, 
$W^{75 \,\, 51}(n)$ is only reliably determined 
for $50 < n < 65$. Elsewhere error estimates indicate that the 
measurement is unreliable, and so only the expansion coefficients in this 
small range are effectively sampled. 
Our measurements at $\beta = 5.71, m=0.1$ with a 
$6^{3}4$ lattice volume of the zeros Fig.(4), and corresponding quark 
number density susceptibility, indicate a first order finite density 
transition at $\mu=0.098(1)$. 

\begin{figure}
\epsfig{file=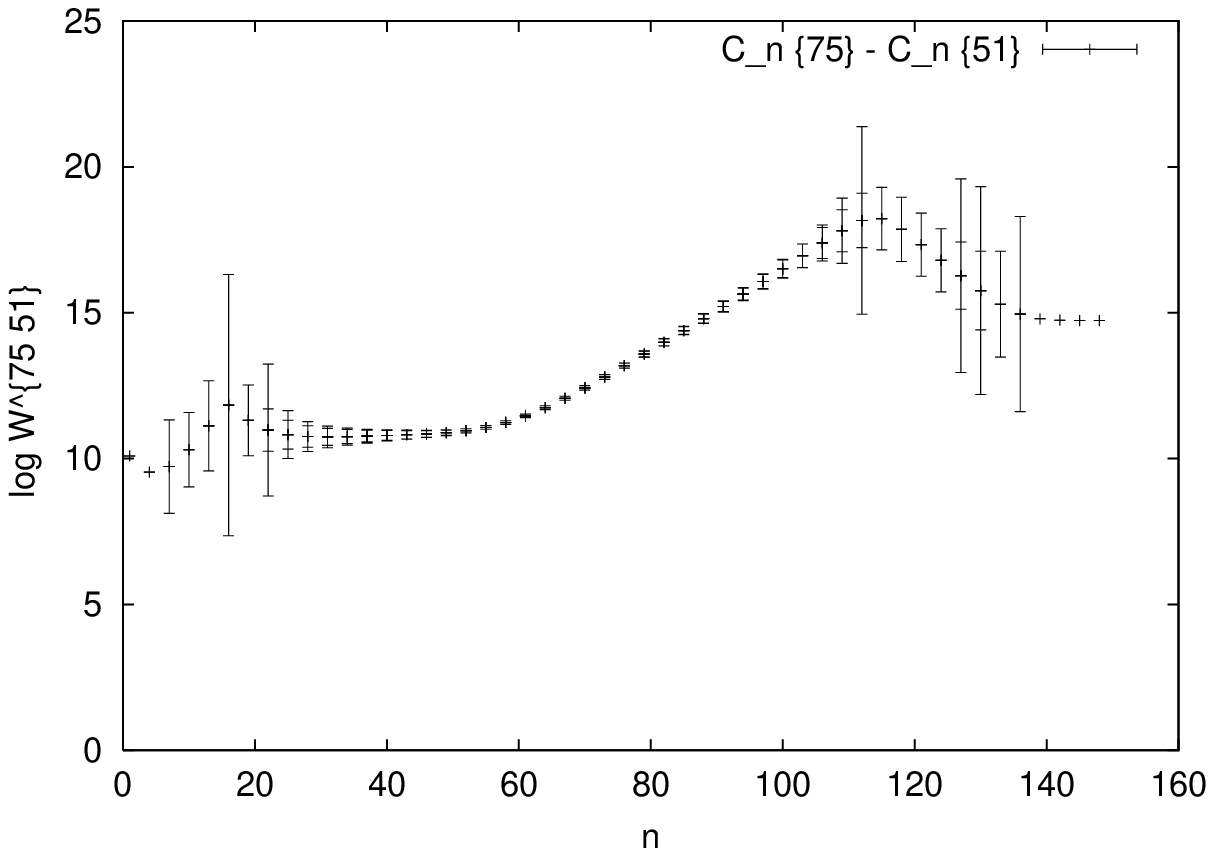, height=0.28\textheight}
\vspace{-1.0cm}
\caption{ Log of the weighting factor ratio for static SU(3) ensembles 
generated using the Monte Carlo measures $c_{75}$ and $c_{51}$, 
defined through 
$W^{75 \,\, 51}(n) =  {\displaystyle{ \left\langle \frac{c_{n}}{c_{75}} 
\right\rangle_{ c_{75} } \left ( \left\langle \frac{c_{n}}{c_{51}} 
\right\rangle_{ c_{51} } \right )^{-1}}}$  .} 
\vspace{1.2cm}
%\label{fig:wr}
\epsfig{file=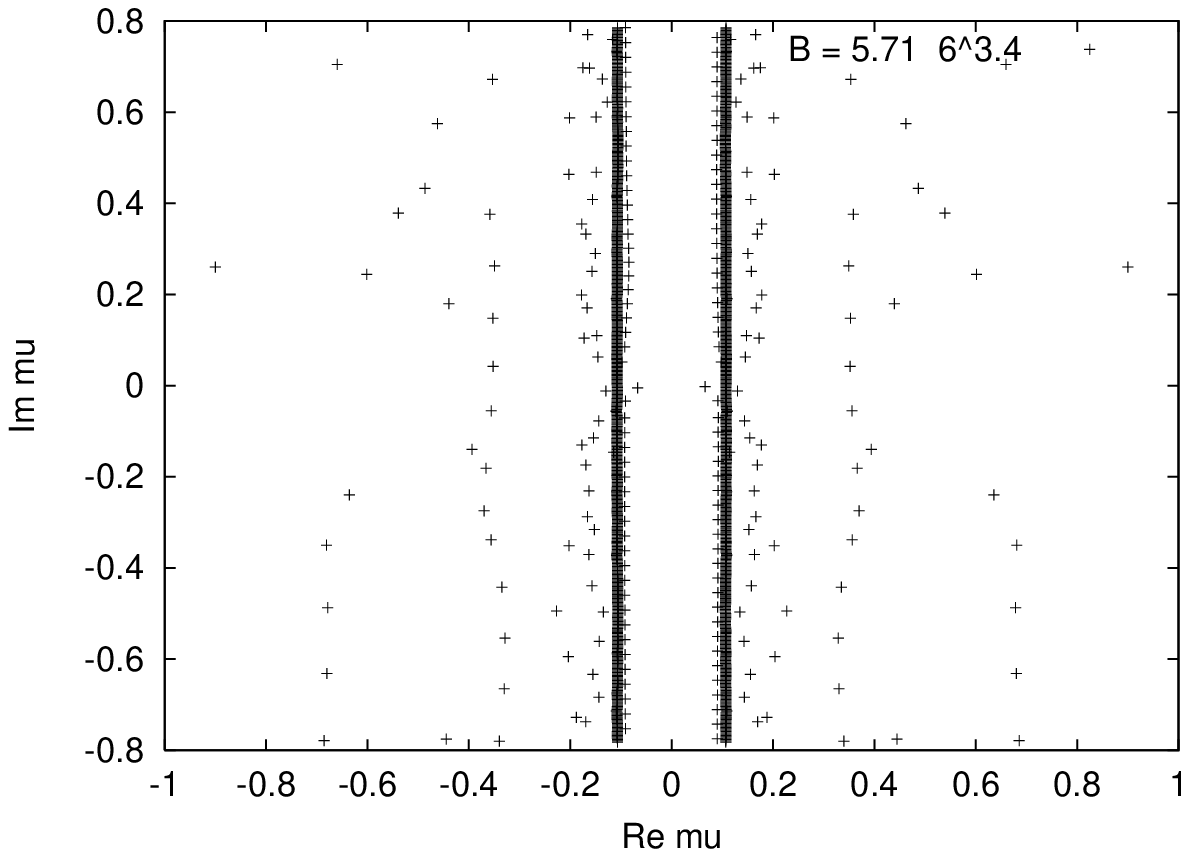, height=0.28\textheight}
\vspace{-1.0cm}
\caption{ Lee Yang zeros evaluated in the complex $\mu$ plane from eleven composite 
weighted ensembles for static SU(3) at intermediate coupling.}
%\label{fig:results}
\end{figure}

\section{Conclusions}
Although we have addressed the pathologies of the Glasgow method with our new 
composite reweighting approach, the main stumbling block to the evaluation of 
finite density SU(3) with dynamical fermions would appear to remain the 
numerical effort required to evaluate the reweighting method. 
We seem to have traded the overlap 
problem for the efficiency our evaluation of canonical ensembles, which we 
believe now requires parallel computing. However, rootfinders can 
often find the zeros of a polynomial from the first few 
coefficients of an expansion, which we can achieve in our method 
by implementing a 
shift variable to truncate the polynomial \cite{10}. Further progress in 
evaluating SU(3) with dynamical quarks at finite density can therefore be made 
by evaluating ensembles weighted with the first few expansion coefficients 
alone in this manner.\\

Thanks to M. Alford, and I. M. Barbour for useful discussions.

%\bibliographystyle{par}
%\bibliography{refs}

\begin{thebibliography}{10}

\bibitem{20}
H. Satz
\newblock {\em Nucl. Phys. Proc. Suppl.} {\bf 94}, 204 (2000).

\bibitem{20a}
H. Heiselberg, V. Pandharipande
\newblock {\em Annu. Rev. Nucl. Part. Sci.} {\bf 50}, 481 (2000).

\bibitem{21}
M. Alford, A. Kaputsin, F. Wilczek
\newblock {\em Phys. Rev. D} {\bf 59}, 502 (1999).

\bibitem{22}
O. Kaczmarek, J. Engels, F.  Karsch, E. Laermann
\newblock {\em Nucl. Phys. Proc. Suppl.} {\bf 83}, 369 (2000).

\bibitem{23}
S. Hands, I. Montvay, M. Oevers, L. Scorzato, J. Skullerud
\newblock {\em Nucl. Phys. Proc. Suppl.} {\bf 457}, 385 (2000).

\bibitem{24}
S. Hands, J.B. Kogut, M. P. Lombardo, S.E. Morrison 
\newblock {\em Nucl. Phys. B} {\bf 558}, 327 (1999).

\bibitem{1}
I.M. Barbour, A.J. Bell
\newblock {\em Nucl. Phys. B} {\bf 372}, 385 (1992).

\bibitem{2}
{J. Kogut {\it et. al.}}
\newblock {\em Nucl. Phys. B} {\bf 225}, 93 (1983).

\bibitem{3}
P.E. Gibbs
\newblock {\em Phys. Lett. B} {\bf 172}, 53 (1986).

\bibitem{4}
C.N. Yang, T.D. Lee 
\newblock {\em Phys. Rev.} {\bf 87}, 404:410 (1952).

\bibitem{5}
C.T.H. Davies, E. Klepfish
\newblock {\em Phys. Lett. B} {\bf 256}, 68 (1991).

\bibitem{6}
I.M. Barbour, J.B. Kogut, S.E. Morrison
\newblock {\em Nucl. Phys. B} {\bf 53}, 456 (1997).

\bibitem{6a}
M.A. Halasz, J.C. Osborn, M.A. Stephanov, J.J.M. Verbaarschot
\newblock {\em Phys. Rev. D} {\bf 61}, 76 (2000).

\bibitem{7}
S. Chandrasekharan, U.J. Wiese
\newblock {\em Phys. Rev. Lett.} {\bf 83}, 3116 (1999).

\bibitem{7a}
S. Hands, J. Kogut, M.P. Lombardo, S.E. Morrison
\newblock {\em Nucl. Phys. B} {\bf 558}, 327 (1999).

\bibitem{7b}
E. Dagotto, F. Karsch, A. Moreo
\newblock {\em Phys. Lett. B} {\bf 169}, 421 (1986).

\bibitem{7c}
J.B. Kogut, D. Toublan, D.K. Sinclair
\newblock {\em hep-lat/0104010}

\bibitem{8}
B. Alles, M. D'Elia, M.P. Lombardo, M. Pepe
\newblock {\em Nucl. Phys. Proc. Suppl.} {\bf 441}, 395 (2001).

\bibitem{8a}
R. Aloisio, V. Azcoiti, G. Di Carlo, A. Galante, A. F. Grillo
\newblock{\em hep-lat/0011079}

\bibitem{8b}
S.J. Hands, J.B. Kogut, S.E. Morrison, D.K. Sinclair
\newblock{\em Nucl. Phys. Proc. Suppl.} {\bf 94} 457 (2001). 

\bibitem{su2}
S. Hands, S. E. Morrison 
\newblock{\em hep-lat/9905021} 

\bibitem{su2a}
S. Morrison, S. Hands
\newblock{\em hep-lat/9902012} 

\bibitem{su2b}
J.B. Kogut, D. Toublan, D. K. Sinclair
\newblock{\em hep-lat/0104010} 

\bibitem{su2c}
M. P. Lombardo
\newblock{\em hep-lat/9907025} 

\bibitem{9}
T. Blum and J. E. Terrick and D. Toussaint
\newblock {\em Phys. Rev. Lett.} {\bf 76}, 1019 (1996).

\bibitem{9a}
O. Kaczmarek, J. Engels, F. Karsch, E. Laermann
\newblock {\em Nucl. Phys. B} {\bf 558}, 307 (1999).

\bibitem{9b}
S. Chandrasekharan
\newblock{\em Nucl. Phys. Proc. Suppl.} {\bf 94} 71 (2001).

\bibitem{10}
I.M. Barbour, R. Buironi, G. Salina
\newblock {\em Phys. Lett. B} {\bf 341}, 355 (1995).

\end{thebibliography}

\end{document}